\documentclass[12pt]{iopart}
\usepackage{iopams}  
\usepackage{graphics}
\usepackage{graphicx}
\usepackage{amssymb}
\usepackage{color}

\newcommand{\TN}{$T_{\rm N}^{4f}$}

\begin{document}

\title{Rare earth magnetism in CeFeAsO: A single crystal study}

\author{A. Jesche$^1$, C. Krellner$^1$, M. de Souza$^2$, M. Lang$^2$ and C.~Geibel$^{1,*}$}
\address{$^1$ Max Planck Institute for Chemical Physics of Solids, D-01187 Dresden, Germany\\
$^2$ Physikalisches Institut, Goethe-Universit\"at Frankfurt, D-60438 Frankfurt(M), Germany}
\ead{$^*$geibel@cpfs.mpg.de}

\begin{abstract}
Single crystals of CeFeAsO, large enough to study the anisotropy of the magnetic properties, were grown by an optimized Sn-flux technique. The high quality of our single crystals is apparent from the highest residual resistivity ratio, $\rm RRR\approx12$, reported among undoped $R$FeAsO compounds ($R=\rm rare$ earth) as well as sharp anomalies in resistivity, specific heat, $C(T)$, and thermal expansion at the different phase transitions. The magnetic susceptibility $\chi(T)$ presents a large easy-plane anisotropy consistent with the lowest crystal electric field doublet having a dominant $\Gamma_6$ character. Curie-Weiss like susceptibilities for magnetic field parallel and perpendicular to the crystallographic $c$-axis do not reveal an influence of a staggered field on the Ce site induced by magnetic ordering of the Fe. Furthermore, the standard signatures for antiferromagnetic order of Ce at $T_N^{4f} = 3.7$\,K observed in $\chi(T)$ and $C(T)$ are incompatible with a Zeeman splitting $\Delta \approx 10$\,K of the CEF ground state doublet at low temperature due to the Fe-magnetic order as previously proposed. Our results can be reconciled with the earlier observation by assuming a comparatively stronger effect of the Ce-Ce exchange leading to a reduction of this Zeeman splitting below 15 K.
\end{abstract}


\today

\submitto{New J. Physics}

\maketitle
Since the discovery of superconductivity at $T_c = 26$\,K in fluorine doped LaFeAsO \cite{Kamihara:2008} and the increase of the superconducting transition temperature $T_c$ up to 55 K in SmFeAsO \cite{ChenNature:2008} the layered FeAs systems and related compounds have been attracting considerable attention. In these compounds superconductivity appears close to the disappearance of an antiferromagnetic (AFM) state upon doping or application of pressure, suggesting an unconventional superconducting state in analogy to those observed in high-$T_c$ or heavy-fermion superconductors. The AFM state below $T_{\rm N}^{\rm Fe}$ in the undoped compounds seems to be of itinerant nature and is connected with a large structural distortion \cite{Cruz:2008, Zhao:2008}. While these unusual features are related to the $3d$ Fe-electrons, the CeFe$Pn$O ($Pn$ = P, As) systems provide in addition strong electronic correlation effects due to the interaction between the Ce $4f$-electrons and the conduction electrons. Thus, CeFePO was shown to be a paramagnetic heavy fermion metal, likely close to a ferromagnetic instability of the Ce $4f$-system \cite{Bruning:2008}, while results on CeFeAsO indicate AFM ordering of Ce in a stable trivalent state \cite{Chen:2008, McGuire:2009}. Recent results prove that the interaction between Fe-magnetism and $4f$-magnetism in CeFeAsO differs strongly from that in the other $R$FeAsO compounds ($R=\rm rare$ earth).
Thus, Chi\,\textit{et al.} \cite{Chi:2008} observed in inelastic neutron measurements on CeFeAsO a splitting of the Kramers doublets at the onset of the Fe-magnetism due to an internal field induced by the AFM Fe-order. 
This is surprising, since the high symmetry of the Ce site, respective to the columnar magnetic Fe structure, should result in the cancellation of Ce-Fe-Heisenberg-type exchange fields. $\mu$SR results \cite{Maeter:2009} evidence a larger hyperfine field at the muon site in CeFeAsO compared to all other $R$FeAsO, which was attributed to an additional contribution originating from the staggered moment on the Ce site induced by the magnetic order of Fe. 
Very recently, investigations of the alloy system CeFeAs$_{1-x}$P$_x$O revealed a switching from an AFM to a ferromagnetic Ce-state at $x \gtrsim 0.4$ \cite{luo:2009}. A likely origin for this unique behavior in CeFeAsO is a stronger hybridization between the $f$-states of Ce and the $3d$-states of Fe near the Fermi level connected with the inherent valence instability of Ce.

Most investigations on the $R$FeAsO systems have been performed on polycrystalline samples. Comparing the reported results with those obtained on $A$Fe$_2$As$_2$ ($A=\rm Ca,$ Sr, Ba, Eu) samples, it is quite obvious that the quality of the $R$FeAsO polycrystals is rather poor. Thus, the residual resistivity ratio (RRR) is usually less than 2 \cite{McGuire:2009}, compared with values extending up to, e.g., 32 in SrFe$_2$As$_2$ systems \cite{Krellner:2008a}. On the other hand, measurements on $R$FeAsO single crystals are presently strongly limited by the very tiny size of the available single crystals, the largest length being typically less than 0.3 mm \cite{Karpinski:2009}. This small size is a consequence of the most widely used preparation technique, based on a single crystal growth in NaCl flux (which has a low solubility for the involved elements) under high pressure, which puts strong limitations on the size of the crucible.

We recently extended our Sn-flux growth technique, which we developed for CeRuPO \cite{Krellner:2008b}, to the growth of CeFeAsO. After some optimization we obtained larger single crystals, which allowed the study of further properties, like e.g., the anisotropy of the susceptibility. 
Further on, both a much higher RRR and narrower anomalies in the specific heat and in the resistivity at the magnetic and structural transitions indicated a much higher quality than the polycrystals reported so far. 
In this paper, we present the investigation and the analysis of the magnetic, thermodynamic, and transport properties of these CeFeAsO single crystals, focused on the magnetism of the Ce $4f$-electrons.

A typical example of a CeFeAsO single crystal is shown in Fig.~\ref{FigM}c, the edges are not very well defined due to the etching procedure after the growth. We found out that CeFeAsO is attacked by diluted HCl on a much higher rate than the CeTPO-compounds (T = Fe, Ru, Os, Co); therefore, we centrifuged the compound in the Sn-matrix before etching for only 10 minutes. The single crystals were characterized with energy dispersive X-ray analysis revealing a stoichiometric Ce:Fe:As content and confirming the presence of oxygen in the compound; furthermore, no foreign phases could be detected. 
We specifically looked at Sn incorporation, but did not observe any Sn line in the microprobe spectra of several 1111 single crystals. In the Sn-flux crystal growth of $A$Fe$_2$As$_2$ compounds, Sn-incorporation was observed to be quite severe for $A$ = Ba, but much weaker for A = Sr, Ca \cite{sun2009}.

Several X-ray powder-diffraction patterns on crushed single crystals recorded on a Stoe diffractometer in transmission mode using monochromated Cu-K$_{\alpha}$ radiation ($\lambda =1.5406$~\AA) confirmed the $P{\rm4/}nmm$ structure type and the formation of single phase CeFeAsO. The lattice parameters $a=4.002(1)$~\AA $ $ and $c=8.647(2)$~\AA $ $, refined by simple least-squares fitting, were found to be in good agreement with the reported structure data \cite{Zhao:2008, Chen:2008, McGuire:2009,Zimmer:1995}. Susceptibility $\chi(T)$ and magnetization $M(H)$  measurements were performed in a commercial Quantum Design (QD) magnetic property measurement system (MPMS). Resistivity  $\rho(T)$ was determined down to 0.4~K using a standard four-probe geometry in a QD physical property measurement system (PPMS). The PPMS was also used to measure the specific heat $C(T)$ with a standard heat-pulse relaxation technique. High-resolution thermal expansion measurements were carried out using a capacitive dilatometer built after \cite{Pott:1983}, enabling to resolve relative length changes $\Delta l/l \geq 10^{-10}$.

\begin{figure}[t]
\begin{center}
\includegraphics[width=12cm]{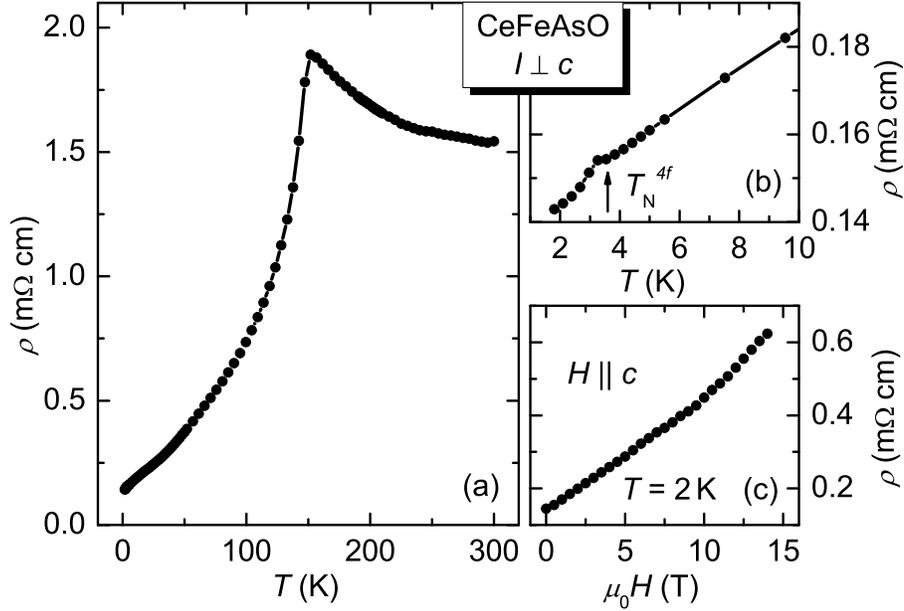}
 \caption{(a) Resistivity as function of temperature for a CeFeAsO single crystal. The sharp drop at $T_{\rm S}^{\rm Fe}, T_{\rm N}^{\rm Fe} \cong150$\,K hallmarks the structural phase transition and the AFM ordering of Fe. (b) $\rho(T)$ around the anomaly at $T_N^{4f}$. (c) The magnetoresistance at $T=2$\,K is linear up to 10\,T and quite large.}
\label{FigRho}
\end{center}
\end{figure}

The quality of our single crystals is demonstrated by the resistivity data presented in Fig.~\ref{FigRho}a. In this  measurement the current was applied along the basal plane. 
The room-temperature value, $\rho_{300\rm K}=1.54$\,m$\Omega$cm, indicates bad metallicity, followed by an upturn towards the structural transition at $T_{\rm S}^{\rm Fe} \cong 149$\,K. 
Below $T_{\rm S} ^ {\rm Fe}$ and $T_{\rm N}^{\rm Fe}$, $\rho(T)$ decreases drastically in a metallic way reaching $\rho_{2\rm K}=140\,\mu\Omega$cm at $T=2$\,K, giving the highest $\rm RRR\approx12$ among the $R$FeAsO samples studied so far.
$T_{\rm S} ^ {\rm Fe}$ and $T_{\rm N}^{\rm Fe}$ were determined from the derivative of the electrical resistivity. 
We are presently performing an analysis of the sample dependence of these transitions, which shall be presented in a forthcoming publication. 
For our best single crystal we found $T_{\rm S} ^ {\rm Fe} = 149$\,K and $T_{\rm N}^{\rm Fe} = 145$\,K.

The high quality of our single crystals is further highlighted by the sharpness of the transitions at $T_{\rm S}^{\rm Fe}$ and $T_{\rm N}^{\rm Fe}$.
The large size of the anomaly in $\rho(T)$, $\alpha(T)$, and $C(T)$ (not shown) confirms the presence of the AFM ordering of Fe in our single crystals.
Furthermore, the pronounced drop in $\rho(T)$ we observe at $T_{\rm S} ^ {\rm Fe}$ and $T_{\rm N}^{\rm Fe}$ supports the absence of any significant Sn-incorporation in our single crystals, because Sn incorporation in Sn-flux-grown BaFe$_2$As$_2$ single crystals results in an increase in resistivity when cooling below $T_{\rm N}^{\rm Fe}, T_{\rm S}^{\rm Fe}$ in contrast to a decrease for samples grown from self-flux (see e.g.\,\cite{sun2009}) and for polycrystalline samples; also recent results show an increase in resistivity when cooling below $T_{\rm N}^{\rm Fe},T_{\rm S}^{\rm Fe}$ in some large LaFeAsO single crystals grown from NaAs-flux, this increase being attributed to defects\,\cite{jan2009}. 
In Fig.~\ref{FigRho}b, $\rho(T)$ is enlarged for $T<10$\,K, where a distinct anomaly is visible at the AFM ordering temperature of Ce, $T_{\rm N}^{4f}$. 
At this anomaly, $\rho(T)$ tends to slightly increase before decreasing with lower $T$. The former suggest the opening of a small gap at the Fermi energy, the latter is because of reduced spin-spin scattering in the AFM ordered state and typical for Ce-systems. Preliminary measurements of the magneto-resistance $\rho(H)$ for magnetic field applied along the $c$ direction, shown in Fig.~\ref{FigRho}c, reveal a huge positive magneto-resistance, reaching $\rho_{\mu_0 H = 14\,\rm T}$/$\rho_{\mu_0H = 0}  = 4.3$. 
For fields up to $\mu_0H = 10$\,T, $\rho(H)$ is linear followed by a steeper increase at higher fields. Further measurements with different magnetic field directions are necessary to clarify the origin of this greatly enhanced magnetoresistance and its peculiar field dependence in CeFeAsO.

\begin{figure}[t]
\begin{center}
\includegraphics[width=12cm]{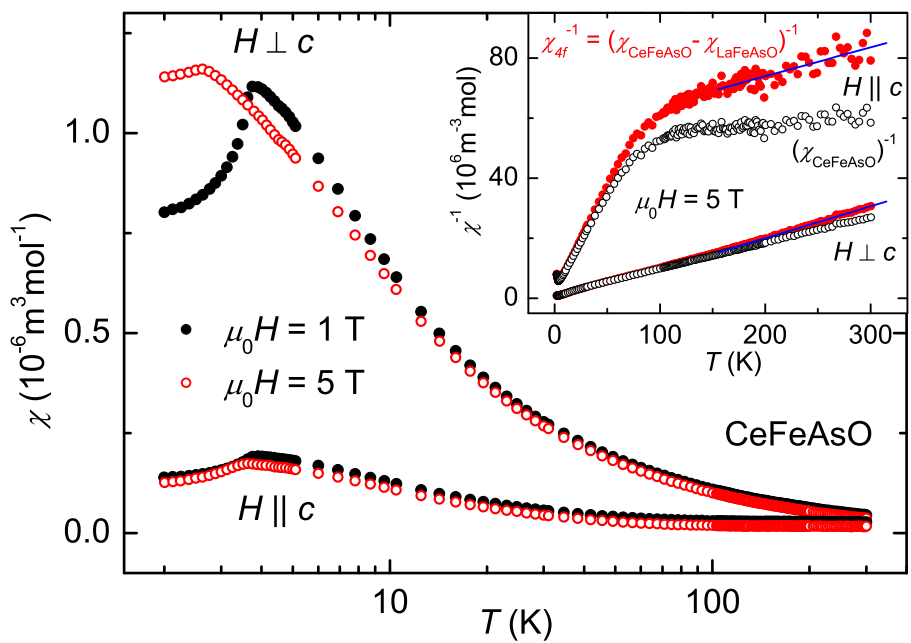}
 \caption{Anisotropic susceptibility of CeFeAsO for $H\perp c$ and $H\parallel c$. The AFM order of the Ce-ions at $T_N^{4f}=3.7$\,K is clearly visible for both crystallographic directions.   Inset: Inverse susceptibility as function of temperature measured in a magnetic field of $\mu_0H = 5$\,T. The open symbols (black) represent the measured data and the closed symbols (red) denotes the $4f$-part of the inverse susceptibility obtained by subtracting $\chi_{\rm LaFeAsO}$ reported in Ref.~\cite{Klingeler:2008}. A Curie-Weiss fit above 150\,K gives $\mu^{\perp}_{\rm eff}=2.45\,\mu_B$ for $H\perp c$ and $\mu^{\parallel}_{\rm eff}=2.6\,\mu_B$ for $H\parallel c$ (blue lines).}
\label{FigChi}
\end{center}
\end{figure}

The magnetic susceptibility of CeFeAsO is dominated by the contribution of the well localized, trivalent Ce. While the paramagnetic susceptibility of Fe-systems is generally very isotropic due to a small spin orbit coupling (SOC), the large SOC of rare earth elements can lead to a huge anisotropy of the magnetic properties. That is the case for CeFeAsO as evidenced by our susceptibility data on a single crystal. 
In Fig.~\ref{FigChi} we compare $\chi(T)$ for $H \perp c$ and $H \parallel c$.  $\chi_{\perp}$ increases continuously below 300 K and reaches a value of 1.16$\cdot$10$^{-6}$m$^3$/mol at the maximum at $T_N^{4f}$. 
In contrast, $\chi_{\parallel}$ is almost constant between 300 and 100 K. Although it starts to increase significantly below 100 K, it reaches only a value of 0.2$\cdot$10$^{-6}$m$^3$/mol at \TN, a factor of 6 smaller than within the basal plane. 
The Curie-Weiss plot $\chi(T)^{-1}$ versus $T$ allows for a more detailed analysis (inset of Fig.~\ref{FigChi}). 
While $\chi_\perp^{-1}$ shows an almost straight line from \TN up to the highest temperatures with an effective moment  $\mu_{\rm eff}^{\rm CeFeAsO} = 2.71\,\mu_B$, slightly larger than that expected for a free Ce$^{3+}$ ion, $\chi_\parallel^{-1}$ increases linearly only up to 80 K, and then stays almost constant. 
For a more precise determination of the $4f$-contribution we subtracted the reported susceptibility of LaFeAsO \cite{Klingeler:2008}, $\chi_{\rm LaFeAsO}$, since the Fe and valence electron contributions are not expected to differ significantly between the La- and the Ce-based compounds. 
Fits (blue lines) to these 
$\chi_{4f}^{-1} = (\chi_{\rm CeFeAsO}  -  \chi_{\rm LaFeAsO})^{-1}$ 
data (closed red symbols in the inset of Fig.~\ref{FigChi}) give an effective moment of
$\mu^{\perp}_{\rm eff}=2.45\,\mu_B$ and a Weiss temperature of $\Theta^\perp = 11$\,K for $H \perp c$ (fit above 150 K), while for $H \parallel c$ we get $\mu^{\parallel}_{{\rm eff-high}T}= 2.6\,\mu_B$ and $\Theta^\parallel_{{\rm high}T} = -580$\,K for the fit above 150 K but $\mu^{\parallel}_{{\rm eff-low}T}=0.96\,\mu_B$ and $\Theta^\parallel_{{\rm low}T} = -3.6$\,K for the fit below 70 K.

The anisotropy of the susceptibility in rare earth compounds is governed by the crystal electric field (CEF) and to a less extent by the anisotropy of the exchange interactions respective to the different components of the local moment. 
The strong anisotropy, we observed at high temperatures, point to a huge CEF anisotropy in CeFeAsO; thus, we can in a first approach neglect the exchange anisotropy. Then, the leading CEF parameter $B_2 ^{\rm 0}$ (Stevens coefficient) can be estimated from the difference between $\Theta_\perp~{\rm and}~\Theta_\parallel$, determined at high temperatures (\cite{Boutron:1973,Avila:2004}): 

\begin{equation}
B_2^{\rm 0} = (\Theta_\perp~-~\Theta_\parallel)~\frac{10}{3\cdot(2J-1)\cdot(2J+3)} 
\end{equation}

In our case the corrections for the Fe-contribution in $\chi_\parallel$ lead to a very large uncertainty in $\Theta^\parallel_{{\rm high}T}$, and therefore the value $B_2 ^{\rm 0} = 62$\,K obtained from our data only corresponds to an order of magnitude estimate. 
In this sense it is in reasonable agreement with the value $B_2 ^{\rm 0} = 37$\,K proposed by Chi \textit{et al.} \cite{Chi:2008} on the basis of neutron scattering results. 
The very well defined slope of $\chi_\parallel^{-1}$ versus $T$ below 70 K allows an estimate of the $c$-axis effective moment of the ground state doublet, and thus, a further insight into the CEF scheme. 
For Ce$^{3+}$ in a local tetragonal surrounding with a large, positive, and dominant $B_2 ^{\rm 0}$ CEF parameter, the CEF ground state is expected to be the $\Gamma_6$ doublet ($\mid \pm 1/2 \rangle$). 
This was indeed proposed to be the case in CeRuPO \cite{Krellner:2008b} as well as in CeFeAsO above the ordering temperature of Fe \cite{Chi:2008}. 
$\Gamma_6$ bares a strongly reduced $c$-axis saturation moment of $\mu_z^{\rm sat}= 0.43\,\mu_B$ and a three times larger $xy$-saturation moment $\mu_{xy}^{\rm sat}= 1.29\,\mu_B$. 
Accordingly, one expects at lower temperatures a strongly reduced effective moment along the $c$ direction, of only $\mu_z^{\rm eff}= 0.74\,\mu_B$, while in the basal plane the effective moment  $\mu_{xy}^{\rm eff} = 2.23\,\mu_B$ is only slightly reduced compared to the value for the full Ce$^{\rm 3+}$ state. 
Our experimental results below 70 K confirm such a strong reduction along the $c$ direction and almost no reduction for the basal plane, and therefore strongly support that the CEF ground state doublet has a dominant $\Gamma_6$ character. The orthorhombic distortion at $T_{\rm S}^{\rm Fe} \cong 150$\,K shall induce a slight mixing of the $\mid\pm\frac{3}{2}\rangle$ and $\mid\pm\frac{5}{2}\rangle$ wave functions in the $\Gamma_6$ state, which might be the origin of the slight difference between the $\Gamma_6$ and the observed $\mu_z^{\rm eff}$.
In the tetragonal phase above 150 K, the excited CEF levels corresponding to a $\Gamma_6$ ground state are two $\Gamma_7$ doublets, $\Gamma_7^{(1)} = \rm cos(\alpha)\,\vert\pm3/2\rangle\,+\,\rm sin(\alpha)\,\vert \mp5/2\rangle$  and $\Gamma_7^{(2)} = -\rm sin(\alpha)\,\vert \pm3/2 \rangle\,+\,\rm cos(\alpha)\,\vert\mp5/2 \rangle$.
Chi $et\,al.$ proposed a rather large (negative) mixing angle $\alpha = -25.6^\circ$\,\cite{Chi:2008}. We calculated the susceptibility for such a CEF scheme assuming the excited levels to be at the energies observed in neutron scattering experiments but allowing for different mixing angles $\alpha$ (only the absolute value of $\alpha$ is relevant for $\chi(T)$). 
We found that $\chi_\parallel$ is quite sensitive to $\left| \alpha \right|$: While the absolute value of 1/$\chi_\parallel$ at 300 K is almost independent of $\left| \alpha \right|$ and matches the experimental value nicely, the slope, $\rm d(1/\chi_\parallel)/\rm dT$, at 300 K decreases with increasing $\left| \alpha \right|$ and becomes negative for $\left| \alpha \right|$\,$\gtrsim$\,20$^\circ$. 
Since we observe a positive slope, our susceptibility data suggest a smaller value of $\left| \alpha \right|$ than proposed on the basis of the neutron results, but unfortunately the large uncertainty in the experimental $\chi_\parallel$ at high temperatures does not allow for a precise determination. 
\\
Inelastic neutron scattering \cite{Chi:2008} and $\mu$SR data \cite{Maeter:2009} indicate a splitting of the CEF ground state doublet and a corresponding Ce polarization, attributed to an Fe-induced internal field at the Ce site below $T_{\rm N}^{\rm Fe}$. Our susceptibility data do not reveal any effect of this splitting and polarization, neither in the $T$- nor in the $H$-dependence, which is at first sight surprising. We shall discuss this point in details later on.
\\
Below \TN~=~3.7\,K, both $\chi_\perp(T)$and $\chi_\parallel(T)$ show a pronounced decrease, indicating AFM ordering of Ce. Here again, no evidence for an Fe-induced polarization of Ce is obvious; the $\chi(T)$ dependence looks like in any  standard Ce$^{\rm 3+}$ system ordering antiferromagnetically. 
With increasing applied field, \TN shifts to lower temperatures, more strongly for field in the easy plane than for field along the hard $c$ direction, as expected for a classical antiferromagnet. 

\begin{figure}[t]
\begin{center}
\includegraphics[width=12cm]{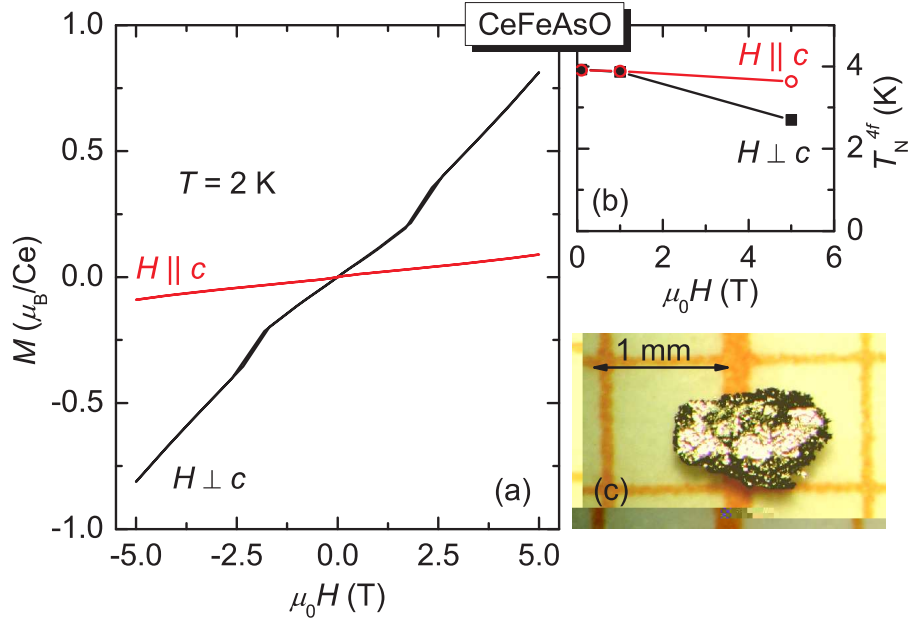}
 \caption{(a) Anisotropic magnetization of CeFeAsO at $T=2$\,K. (b) $T_N^{4f}$ determined from $\chi(T)$ measurements (Fig.~\ref{FigChi}) as function of the external magnetic field. (c) Photograph of a CeFeAsO single crystal, the scale labels 1\,mm.}
\label{FigM}
\end{center}
\end{figure}

A preliminary sketch of the $H-T$ phase diagram is shown in the inset of Fig.\,\ref{FigM}b. In the main part of this figure, Fig.\,\ref{FigM}a, we present the magnetization of a CeFeAsO single crystal at $T=2$\,K, i.e., in the AFM ordered state, for both crystallographic directions. For $H\perp c$, $M(H)$ increases linearly for fields $\mu_0H<2$\,T, followed by a small metamagnetic transition around $\mu_0H=2$\,T. The origin of this metamagnetic transition is presently not settled and thus deserves further magnetization as well as microscopic measurements. 
For $\mu_0H > 2$\,T the magnetization increases again linearly, reaching an absolute value $M=0.8\,\mu_B$/Ce at $\mu_0H=5$\,T, which is about 60\% of the basal plane saturation moment, $\mu_{xy}^{\rm sat}$ expected for a $\Gamma_6$ doublet. 
Assuming a linear extrapolation of $M(H)$ this saturation value would be reached at $\mu_0H_{\rm sat} \approx 7.8$\,T. For $H\parallel c$, $M(H)$ presents a perfectly linear behavior up to $\mu_0H=5$\,T with a much smaller slope $dM/dB = 0.019~ \mu_B/\rm T$. These results are in agreement with the easy-plane anisotropy of CeFeAsO determined by the susceptibility measurements discussed above.
\\
Within a simple molecular field model, defining the molecular field $B_{xy} ^{\rm eff}$ induced by the antiferromagnetic Ce-Ce coupling constant $\lambda_{xy} ^{\rm AFM}$ as
\begin{equation}
 B_{xy} ^{\rm eff} = \lambda_{xy} ^{\rm AFM} \cdot m_{xy}^{\rm Ce},
\end{equation}
 with $m_{\rm xy}^{\rm Ce}$ being the ordered Ce moment (in analogy to equation 12 in \cite{Maeter:2009}), one can estimate $\lambda_{xy} ^{\rm AFM}$ either from the initial slope $dM/dB$ = 0.11 $\mu_B/T$ 
\begin{equation}
\lambda_{xy} ^{\rm AFM} = 1/(2 dM/dB) = 4.5~\rm T/\mu_B
\end{equation}
or from the saturation field $B_{\rm sat}$ 
\begin{equation}
\lambda_{xy} ^{\rm AFM} = B_{\rm sat}/(2 \mu_{xy}^{\rm sat}) = 3.0~\rm T/\mu_B. 
\end{equation}
Within the same approximation, an independent estimation can be obtained from \TN, 
\begin{equation}
\lambda_{xy}^{\rm AFM} = (k_{B}/\mu_B) \cdot $\TN$ / (\mu_{xy}^{\rm sat})^2 = 3.4~\rm T/\mu_B,
\end{equation} 
with $\mu_{xy}^{\rm sat}$ in units of $\mu_{\rm B}$.
All these values are in reasonable agreement, suggesting a rather conventional type of magnetic ordering. 
Accordingly, the initial slope $dM/dB$ for field along the $c$-axis leads to $\lambda_z ^{\rm AFM} =  26~\rm T / \mu_B$. The fact that Ce orders in the basal plane and not along the $c$ direction gives an upper limit for $\lambda_z^{\rm AFM}$ 
\begin{equation}
\lambda_z^{\rm AFM} <  (k_B/\mu_B) \cdot T_N^{4f} / (\mu_z^{\rm sat})^2 = 30~\rm T /\mu_{\rm B}. 
\end{equation}

Comparison with $\lambda^{\rm AFM}_{\rm xy}$ suggests a huge exchange anisotropy, with a much larger exchange for the $z$ component. However, this merely reflects the anisotropy of the $g$ factor, since in the molecular field approximation $\lambda$ is proportional to $g^{-2}$, (e.g. \cite{Blundell:2001}), which is a factor 9 higher for the $z$ direction compared to the $xy$-direction. 
Thus, the product $\lambda_i \cdot g_i ^2$, which is proportional to the exchange parameter in a Heisenberg picture, is slightly lower for the $z$ component than for the basal plane component. 
\begin{figure}[t]
\begin{center}
\includegraphics[width=12cm]{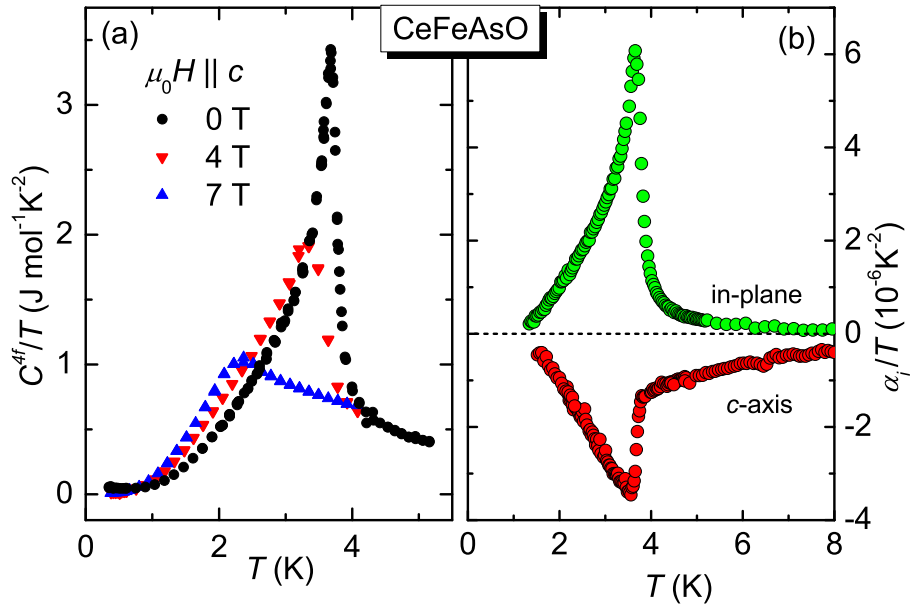}
 \caption{(a) $4f$-part of the specific heat of CeFeAsO around \TN. The anomaly shifts to lower $T$ in an external applied magnetic field. (b) Linear thermal expansion coefficient $\alpha_i$ as $\alpha_i/T$
for single crystalline CeFeAsO measured along an unspecified in-plane (\emph{ab}) direction, $\alpha_{\perp}$, and along the \emph{c}-axis, $\alpha_\parallel$.}
\label{FigHC+TA}
\end{center}
\end{figure}
\\
Next, we focus on the specific heat measurements around \TN, plotted as $C^{4f}/T$ vs $T$ in Fig.~\ref{FigHC+TA}a. The $4f$-contribution, $C^{4f}$, to the specific heat was obtained by subtracting the reported specific heat data of LaFeAsO \cite{Dong:2008} which, however, is small compared to the large anomaly due to the onset of AFM ordering of the Ce-ions.
This sharp $\lambda$-type anomaly reflects the high quality of our single crystal compared to measurements on polycrystals \cite{Chen:2008}. 
The value $C^{4f}_{\rm max} = 12.6$\,J/(mol~K) at the peak at \TN is almost identical to the value 12.5 J/(mol~K) calculated in a simple mean field model for a doublet system. 
From the anomaly in $C^{4f}/T$ the AFM transition temperature can be accurately determined yielding $T_N^{4f}=3.7$\,K. 
The anomaly shifts to lower temperatures in an applied magnetic field, similar to what is observed in the susceptibility measurements (see Fig.~\ref{FigChi} and \ref{FigM}b). 
In zero field, $C^{4f}/T$ decreases quadratically with temperature below $T_N^{4f}$, supporting that the excited magnons are of AFM nature. 
Below $T~=~0.8$\,K, $C^{4f}/T$ levels off at a constant Sommerfeld-coefficient, $\gamma_0=50$\,mJ(mol K$^2)^{-1}$, which is enhanced by more than one order of magnitude compared to LaFeAsO ($\gamma_0^{\rm La}=3.7$\,mJ(mol K$^2)^{-1}$ \cite{Dong:2008}). 
This enhancement can therefore be attributed to correlations of the $4f$-electrons. 
Whether or not these correlations are also entangled to the $3d$-electrons can presently not be answered and deserves further more microscopic studies. 

\begin{figure}
\begin{center}
\includegraphics[width=12cm]{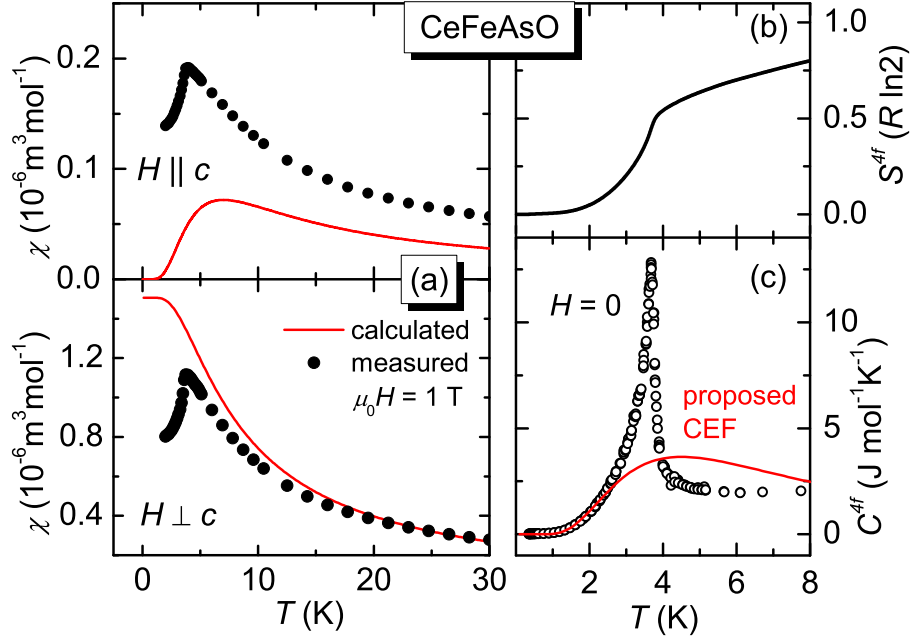}
\caption{(a) Calculated (see text) and measured magnetic susceptibility for temperatures below 30 K. (b) Magnetic entropy calculated by integrating $C^{4f}/T$ over $T$. (c) Specific heat plotted as $C^{4f}$ vs $T$; for comparison the specific heat calculated from the proposed CEF scheme \cite{Chi:2008} is drawn as a line (see text).}
\label{chi+S}
\end{center}
\end{figure}

The magnetic entropy was obtained by integrating $C^{4f}/T$ over $T$ and is presented in Fig.~\ref{chi+S}b. 
The steep increase at $T_{\rm N}^{4f}$ together with absolute values of 80\% of $R\ln 2$ at $T=8$\,K confirm that most of the $4f$-degrees of freedom of the lowest CEF doublet are involved in the AFM order.
\\
Finally, in Fig.\,\ref{FigHC+TA}b we show the results of the uniaxial thermal expansion coefficient $\alpha(\textit{T}) = \textit{l}^{-1}(\partial \textit{l}/\partial \textit{T})$ along a non-specified direction within the $ab$-plane, $\alpha_{\perp}$ and along the $c$-axis, $\alpha_{\parallel}$, for CeFeAsO below 8\,K in a representation $\alpha/T$ vs. $T$. 
The data are dominated by pronounced and strongly anisotropic phase transition anomalies at the AFM ordering of the 4$f$-moments at $T^{4f}_N$ = (3.7 $\pm$ 0.1)\,K. 
Both the sharpness of the transition and the pronounced anisotropy confirm the high quality of the single crystal studied here. 
As shown in Fig.\,\ref{FigHC+TA}b, the negative discontinuity in $\alpha_{\parallel}$ is preceded by an anomalous negative expansivity, unexpected for a lattice contribution, but likely to originate from the Ce-4$f$ degrees of freedom. 
Its peculiar $T$-dependence suggests a relation to the decrease of the Zeemann splitting discussed later on. 
The $c$-axis behavior contrasts with a small and positive $\alpha_{\perp}$ revealed above $T^{4f}_N$. 
Note that $\alpha_{\perp}$ can be significantly affected by the domain structure which forms upon cooling through the tetragonal to orthorhombic phase transition at $T_{S}^{\rm Fe} \cong 150$\,K. 
Details of the expansivity results at $T_{\rm S}^{\rm Fe}$ will be published elsewhere.

Combining thermal expansion with specific heat data (Fig.\,\ref{FigHC+TA}) and making use of the Ehrenfest relation, 
\begin{equation}
\frac{dT^{4f}_{N}}{dP_{i}} = V_{mol} \cdot T^{4f}_{N} \cdot \Delta \alpha_i / \Delta C,
\end{equation}

an estimate of the pressure dependence of $T^{4f}_{N}$ can be made. Using the molar volume $V_{mol}$ = (4.17 $\cdot$ 10$^{-5}$)\,m$^3$/mol, $\Delta \alpha_\parallel$ = $-$(10.5 $\pm$ 0.5) $ \cdot$ 10$^{-6}$ \,K$^{-1}$ estimated from Fig.\,\ref{FigHC+TA}b and $\Delta C^{4f}$ = 12.6\,Jmol$^{-1}$K$^{-1}$, one finds $dT^{4f}_{N}/dP_{\parallel}$ = $-$0.013\,K/kbar.
Although a quantitative estimate of the in-plane pressure dependence is precluded by potential domain structure effects, the large positive discontinuity in $\alpha_{\perp}$ at $T^{4f}_N$ in Fig.\,\ref{FigHC+TA}b indicates a positive in-plane pressure effect on $T^{4f}_{N}$. 
Since a very small pressure dependence of $T^{4f}_{N}$ is expected under hydrostatic pressure conditions, 
(substituting large As by smaller P left \TN almost unaffected up to $x \lesssim 0.3$ in CeFeAs$_{1-x}$P$_x$O \cite{luo:2009}), a significant in-plane anisotropy in $dT^{4f}_{N}/dP_{a,b}$ can be inferred.
\\
According to the results and discussion presented so far, CeFeAsO behaves like a standard trivalent Ce$^{3+}$ system ordering antiferromagnetically at \TN~=~3.7\,K, with a strong easy-plane CEF anisotropy. 
There is neither evidence for a large Kondo effect as in CeFePO, nor direct evidence for an influence of the SDW ordering of Fe. 
More specifically, the observation of a well-defined AFM transition at \TN associated with a large entropy seems to be in strong contradiction with a Zeeman splitting of the CEF ground state of 0.93 meV = 10.8 K as proposed by Chi\,$et~al.$ \cite{Chi:2008}, on the basis of inelastic neutron scattering measurements. 
On the one hand, for such a splitting the polarization of the ground state already amounts to 92\,\% at 4.5 K, i.e., the system would already be almost completely polarized at a temperature 20\% above the observed \TN. Thus, there would not be much degrees of freedom left for a magnetic transition. 
Accordingly, most of the entropy of $R$ln2 associated with the CEF ground state doublet would have been released at higher temperatures, the entropy at 4 K left for a transition would amount to only 34\,\% of $R$ln2, far below the observed value 55\,\% of $R$ln2. 
Furthermore, the $4f$ specific heat should be quite different from that observed experimentally. We have calculated the specific heat using the appropriate Schottky-model for 4 non-degenerated energy levels at $E$\,=\,0,\,0.93,\,16.9,\,20.1\,meV as proposed by Chi \textit{et al.} \cite{Chi:2008}. 
In Fig.~\ref{chi+S}c the resulting specific heat curve is compared to our experimental data as $C^{4f}$ vs. $T$ around $T_N^{4f}$. The calculated curve shows the expected broad Schottky-type anomaly with maximal values $C_{\rm max}=3.65$\,J/molK at $T=4.5$\,K. 
This is in strong contrast to the $\lambda$-type anomaly with absolute values $C^{4f}=12.6$\,J/molK at $T_N^{4f}$ and the almost $T$-independent $C(T)$ above 5 K.
\\
In order to shed more light into this apparent contradiction, we performed a more profound analysis of our results. The first point we looked at in more detail was how the Ce susceptibility is affected by an internal field due to the SDW ordering of Fe. 
We assumed a $\Gamma_6$ CEF ground state doublet, which corresponds to an effective $S = \frac{1}{2}$ state with an anisotropic $g$ factor $g_{xy} = 2.57$ and $g_z = 0.86$, under the influence of both an internal field $B_{\rm int}$ and an external field $B_{\rm ext}$. Maeter $et~al$ \cite{Maeter:2009}, give strong evidence that the internal field on the Ce site due to the Fe ordering is a staggered field parallel to the $c$-axis. We followed this proposition and took $B_z ^{\rm int} = 18.75$\,T as required for a splitting 
\begin{equation}
\Delta = B_z ^{\rm int} \cdot g_z \cdot \mu_B/k_B = 10.8\,K. 
\end{equation}
We assumed $B_z ^{\rm int}$ to be $T$-independent since we were interested in the low-temperature behavior ($T < 30$\,K), where the Fe-moment is already saturated. 
For an external field along the $c$ direction, the magnetization and the susceptibility can then easily be calculated using the usual Brillouin function $B_{1/2}(x)$ for two different Ce-sites, one with $B_{\rm ext}$ and $B_{\rm int}$ parallel and one with $B_{\rm ext}$ and $B_{\rm int}$ antiparallel. We found that $B_{\rm int}$ has almost no effect above 15 K, but below 15 K it starts to reduce $\chi_\parallel$, such that the $c$-axis susceptibility should show a maximum at $T = 7$\, K followed by a pronounced drop below 6\,K (Fig.\,\ref{chi+S}a, upper part). 
This can easily be understood from the Brillouin function since the susceptibility is now related to the slope at 
\begin{equation}
x = \pm B_{\rm int} \cdot g_z \cdot \mu_B/(2 k_B T)
\end{equation}
 instead of the slope at the origin $x = 0$. As long as $x$ is in the linear part of the Brillouin function, there is no difference, but once one reaches the saturation region the slope vanishes. We also studied $\chi_\perp$; 
for a Zeeman splitted doublet due to $B_{\rm int}$ along the $c$-axis, $\chi_\perp$ corresponds to a Van Vleck susceptibility due to mixing between the $\mid+1/2\rangle$ and the $\mid-1/2\rangle$ states:
\begin{equation}
\chi_\perp = \frac{N_{\rm A}\,\mu_0\,\mu_B\,g_{xy}^2}{2\,g_z\,B_{\rm int}}\,{\rm tanh}\lbrace \frac{\Delta}{2k_{{\rm B}}T}\rbrace
\end{equation}

This leads to a constant $\chi_\perp$ at low $T$, but as soon as the temperature gets closer to $\Delta/k_B$, the excited state gets populated and $\chi_\perp$ merges in a Curie law with the usual moment (Fig.\,\ref{chi+S}a, lower part).  
The calculation revealed that the departure from the Curie law is only significant below 5 K, thus in a temperature range where the experimental data are affected by the antiferromagnetic ordering. 
In conclusion, this analysis clarifies the absence of any effect of $B_{\rm int}$ on $\chi_\perp(T)$ above \TN and on $\chi_\parallel(T)$ above 15 K, but confirms that the observed continuous increase of $\chi_\parallel(T)$ below 10 K down to $T_N^{4f}$ is in clear disagreement with the suggested splitting, which should lead to a visible maximum. 
This cannot be repaired by changing the orientation of $B_{\rm int}$, since then an equivalent maximum would appear in $\chi_\perp(T)$. Therefore, both the specific heat and $\chi_\parallel$ contradict a splitting of the CEF ground state of the order of 11 K below $T < 10$\,K. 

\begin{figure}
\begin{center}
\includegraphics[width=12cm]{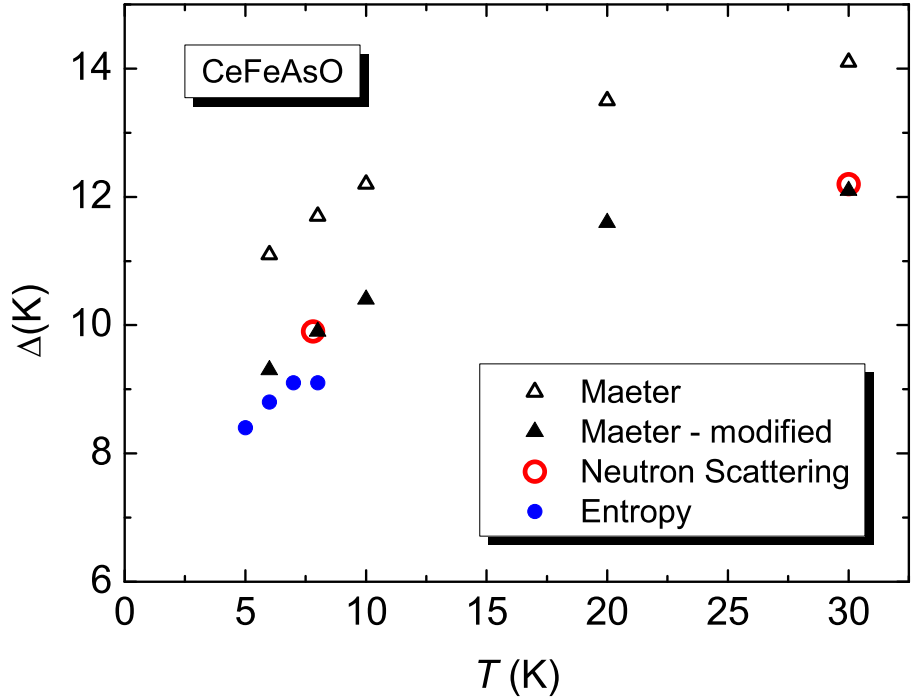}
\caption{Gap energy $\Delta$ as a function of temperature estimated from neutron scattering \cite{Chi:2008} and entropy data. The open triangles are the values calculated with the model and parameters of Maeter $et~al.$ \cite{Maeter:2009}. In a modified calculation (closed triangles) the Ce-Fe exchange field was slightly reduced to reconcile experimental and calculated values.}
\label{delta}
\end{center}
\end{figure}

Thus, we are left with a contradiction between the clear evidence for a splitting at higher temperatures in the neutron and $\mu$SR results and the absence of the expected effects of this splitting in $\chi_\parallel$ and $C(T)$ below 10 K. 
This contradiction can be resolved by assuming a reduction of the splitting towards lower temperatures. A closer inspection of the neutron data of Chi $et~al.$ \cite{Chi:2008} supports such an assumption. 
The low-energy spectra in the inset of Fig.\,2a of this reference reveal a quite sharp excitation at 1.05 meV (= 12.2 K) for $T = 30$\,K and $T = 96$\,K, but for $T = 7.8$\,K the peak has shifted to 0.85 meV (= 9.9 K), and seems to have broadened on the low-energy side. 
In order to get a further idea about the possible magnitude of a $T$-dependent splitting, we assumed that above 5 K the system can be fully described by two singlets separated by a $T$-dependent splitting $\Delta(T)$, and calculated $\Delta(T)$ from the measured entropy $S^{4f}$,
\begin{equation}
 S^{4f} = R(p_0 {\rm ln}p_0 + p_1 {\rm ln}p_1) ~~~
 {\rm with}~~~ p_1/p_0 = {\rm exp}\lbrace \frac{-\Delta(T)}{k_{\rm B} T} \rbrace.
\end{equation}
The results are plotted in Fig.\,\ref{delta} together with the peak positions in the low-energy inelastic neutron data. Both data sets consistently indicate a decrease of the splitting, especially below 10 K.
A $T$ dependent internal field was already introduced by Maeter $et~al.$ \cite{Maeter:2009}. In order to get a better fit to their $\mu$SR data, they considered a further exchange field on the Ce-site due to the polarization of the Ce itself:
\begin{equation}
B_z ^{\rm int} = B_z ^ {\rm Fe-Ce} + J_{3z} \cdot m_z^{\rm Ce}
\end{equation}
where the exchange field due to the Fe was assumed to be $T$-independent below 30 K, $B_z ^{\rm Fe-Ce} = B_0$, $J_{3z}$ corresponds to a Ce-Ce coupling constant for a specific symmetry of the Ce exchange, and $m_z ^{\rm Ce}$ is the polarization of the Ce-site.

Maeter $et~al.$ \cite{Maeter:2009} obtained the best agreement with the $\mu$SR data for $B_0 = 26.8$\,T and $J_{3z} = -24.3\,\rm T /\mu_B$. Interestingly, this coupling parameter is quite close to the coupling constant we estimated from $\chi_\parallel$ below $T_{\rm N}$ (see discussion of susceptibility data). 
The negative sign of $J_{\rm 3z}$ leads to a reduction of the total internal field with decreasing temperature, because of the increasing polarization of the Ce in the Fe exchange field. Accordingly, the splitting of the $\Gamma_6$ ground state becomes $T$ dependent. 
\\
We calculated $\Delta(T)$ using the parameter of Maeter $et~al.$ \cite{Maeter:2009} ($m_z ^{\rm Ce}$ was calculated in an iterative process using the Brillouin function), and plotted the results in Fig.\,\ref{delta}, where they can be compared with the values deduced from neutron and specific heat experiments as described above. 
While the calculated $T$-dependence shows the expected decrease of $\Delta(T)$ with decreasing $T$, the absolute values are significantly above the values we deduced from experiments. A nice agreement can be obtained by just reducing the Fe-Ce-exchange field to $B_0 = 23$\,T. 
The effect of the Ce-Ce exchange is then comparatively stronger, and is already quite sizeable at 10 K. The broadening on the low-energy side of the excitation in the low-energy neutron spectra at $T = 7.8$\,K, as well as further analysis of the $T$ dependence of the specific heat and of the susceptibility indicate that the effect of the Ce-Ce exchange increases more strongly towards lower $T$ than suggested by this simple model. 
One would e.g. expect that the inherent tendency of Ce towards an own magnetic ordered state leads to short-range correlations above the ordering temperature, which would be associated with a fluctuating transverse field, not accounted for in the simple 'static' model. 
In summary, our analysis of the susceptibility and specific heat data together with a reassessment of the neutron data indicate a strong effect of the Ce-Ce exchange at lower temperatures, competing with the exchange field induced by the Fe-ordering, and leading to a decrease of the total internal field at the Ce site starting already below around 15 K.
\\
In conclusion, we have grown larger single crystals of CeFeAsO using a Sn-flux technique and investigated the magnetic properties related to Ce by means of $\chi(T)$, $M(H)$, $C(T,H)$,  $\rho(T,H)$, and $\alpha(T)$ measurements. 
In contrast to earlier results on polycrystals, the basal plane resistivity in our single crystals not only shows a sharp drop at the structural (and the antiferromagnetic) transition at $T \cong 150$\,K, but continues to decrease down to the lowest temperature, resulting in a high $\rm RRR \approx 12$. 
This observations evidence on the one hand the presence of a well-defined metallic state in CeFeAsO at low temperatures, and on the other hand the high quality of our single crystals, which is also demonstrated by the sharpness of the anomalies in the resistivity, specific heat, and thermal expansion at the different transitions. 
The magnetic susceptibility, which is dominated by the contribution of the well localized $4f$ electrons of Ce$^{3+}$, presents a strong easy-plane anisotropy, with predominantly $\Gamma_6$ ($\mid\pm1/2\rangle$) character.
At \TN~=~3.7\,K, maxima in the susceptibility along both directions as well as a sharp $\lambda$-type peak in the specific heat and in the thermal expansion evidence a transition to an antiferromagnetic state. 
The large amount of entropy released at the transition implies that most of the degrees of freedom of the CEF ground state are involved in the formation of the antiferromagnetic state. 
This is also confirmed by the peak value of $C(T)$ being very close to that expected for a mean-field transition. This large amount of entropy at \TN as well as the continuous increase of $\chi_\parallel$ below 15\,K down to \TN are incompatible with a $T$-independent Zeeman splitting of the CEF ground state $\Delta = 10.8$\,K as suggested from neutron measurements. 
The analysis of the specific heat and reassessment of the neutron data indicate a significant reduction of the splitting below 15 K. 
This reduction can certainly be attributed to the growing effect of Ce-Ce exchange, which is competing with the Fe-Ce exchange and thus leads to a reduction of the molecular-field at the Ce-site. 
Although this is likely a precursor effect of the Ce-ordering at \TN~=~3.7\,K, it starts to become visible at a much higher temperature, around 15 K. 
The $4f$-electrons in the ordered state show pronounced correlations, reflected in an enhanced $\gamma_0=50$\,mJ/molK$^2$, more than one order of magnitude higher than in the closely related compound LaFeAsO without $4f$-electrons. 
The observation of a heavy fermion state connected with strong ferromagnetic Ce-Ce exchange in CeFePO suggests a very complex interplay between $4f$ and $3d$ physics in the alloy CeFeAs$_{1-x}$P$_x$O, both on the level of exchange effects and on the level of strong correlation effects. 
This certainly deserves further investigations. 
\\
The authors thank P. Scheppan and U. Burkhardt for energy dispersive X-ray analysis of the
samples and N. Caroca-Canales and R.Weise for technical assistance in sample
preparation. H. Rosner, Q. Si, and M. Nicklas are acknowledged for
valuable discussions.
\\

\end{document}